\documentclass[prd,nofootinbib,twocolumn,superscriptaddress,preprintnumbers,balancelastpage]{revtex4-1}

\pdfoutput=1
\usepackage{amsmath,amssymb}
\usepackage{amsfonts}
\usepackage{bm,bbm}
\usepackage{graphicx}
\usepackage{mathrsfs}
\usepackage{slashed}
\usepackage{booktabs}
\usepackage{tabu}
\usepackage[dvipsnames]{xcolor}
\usepackage[normalem]{ulem}
\usepackage{tikz}
\usepackage{soul}
\usepackage{extarrows}

\usepackage{hyperref}
\hypersetup{colorlinks,citecolor= blue,linkcolor= blue, urlcolor=blue}

\usepackage{xcolor}
\usepackage{ulem}
\usepackage{array}
\usepackage{verbatim}
\usepackage{epsfig}
\usepackage{multirow}
\usepackage{physics}
\usepackage{mathdots}
\usepackage{yhmath}
\usepackage{cancel}
\usepackage{siunitx}
\usepackage{amssymb}
\usepackage{gensymb}
\usepackage{tabularx}
\usepackage{extarrows}
\usepackage{booktabs}
\usetikzlibrary{fadings}
\usetikzlibrary{patterns}
\usetikzlibrary{shadows.blur}
\usetikzlibrary{shapes}
\tikzset{every picture/.style={line width=0.75pt}} 

\newcommand{\be}{\begin{equation}}
\newcommand{\ee}{\end{equation}}
\newcommand{\ba}{\begin{array}}
\newcommand{\ea}{\end{array}}
\newcommand{\bea}{\begin{eqnarray}}
\newcommand{\eea}{\end{eqnarray}}

\usepackage{ulem,fancyvrb}
\usepackage{xcolor} 

 \newcommand{\zb}[1]{{\color{black} {#1}}}

\newcommand{\vn}{\vec{n}}

\begin{document}

\title{Probing \zb{a} muonic force with \zb{the} periastron advance in binary pulsar systems}

\author{Zuowei Liu}
\thanks{Corresponding author.\\ zuoweiliu@nju.edu.cn}
\affiliation{Department of Physics, Nanjing University, Nanjing 210093, China}

\author{Zi-Wei Tang}
\affiliation{Department of Physics, Nanjing University, Nanjing 210093, China}

\begin{abstract}
Pulsars, highly magnetized, rotating neutron stars, can have significant muon abundances in their dense cores, making them promising environments to probe ultralight mediators coupled to muons. The precise measurement of periastron advance in binary pulsar systems provides a sensitive probe of such long-range forces. In this work, we study the periastron advance constraints from binary pulsar systems on the ultralight muonic mediators. We compute the muon number fraction in neutron stars, by properly taking into account the suppression effect of the long-range muonic force. We find that the periastron advance constraints impose the most stringent constraints on ultralight muonic mediators in the mass range of $\simeq(10^{-17},\,2\times10^{-15})$ eV, probing muonic couplings as small as $\mathcal{O}(10^{-21})$, which surpass the limits from LIGO/Virgo gravitational wave measurements, by about an order of magnitude. 
\end{abstract}

\maketitle

\section{Introduction}\label{sec:intro}

The exploration of new long-range
forces has spanned a broad range 
of distances and coupling strengths \cite{Will:2014kxa}. 
The long-range forces that 
act on neutrons, protons, or electrons 
are tightly constrained by the experimental tests of
the weak equivalence principle (WEP)
\cite{Tino:2020nla, Adelberger:2006dh,Schlamminger:2007ht,Wagner:2012ui,Berge:2017ovy, Williams:2004qba,Turyshev:2006gm}. 
On the other hand, the long-range forces that act on muons  
remain largely unconstrained, 
because muons are rare in most WEP experiments. 
One of the ways to search for muonic
forces is to study their effects on the inspiral
dynamics of binary systems involving neutron stars (NS) 
that have significant muon fractions \cite{Haensel:2007yy}. 
Constraints on muonic forces from such systems 
have been carried out through gravitational wave signals  
\cite{Dror:2019uea,Xu:2020qek} 
and orbital decays 
\cite{Dror:2019uea,KumarPoddar:2019ceq,Cheng:2023qys}.

In this work, we use periastron advance of the 
binary pulsar system to search for 
ultralight mediators coupled to muons. 
The periastron for a binary star system marks the point of closest orbital approach \cite{Poisson_Will_2014}. Periastron advance, then, describes the secular precessional motion of this point.\footnote{The phenomenon ``periastron advance'' is conceptually identical to the perihelion precession observed in the solar system, while ``perihelion'' is used specifically for the closest approach to the Sun \cite{Poisson_Will_2014}.}
The periastron advance for many binary pulsar
systems {has} been measured with great {precision}
\cite{Kramer:2021jcw, Weisberg:2016jye, 
vanLeeuwen:2014sca,Ferdman:2014rna,Jacoby:2006dy,Fonseca:2014qla}. 
Recently, tests of general relativity (GR) and beyond
\cite{Stairs:2003eg,Will:2014kxa,Wex:2014nva,Berti:2015itd,Freire:2024adf}, 
as well as searches for beyond-the-Standard-Model
(BSM) physics
\cite{Blas:2019hxz, Kus:2024vpa,Fabbrichesi:2019ema} 
using the precise pulsar periastron advance measurements 
have been carried out.
Because the periastron advance induced by muonic forces 
is strongly dependent on the muon abundance in the neutron star core,  
it is of great importance to properly take into account 
the effects of the muonic Yukawa potential, 
which can significantly decrease the muon abundance. 
This suppression arises from the potential energy
associated with the muonic force, which raises
the energy needed to produce muons in the NS
core, thereby reducing the muon fraction. 
Although such an effect has been mentioned in
Ref.~\cite{Dror:2019uea}, 
it has not been quantitatively analyzed for 
muon abundances in neutron star cores before. 
We find that the suppression of the NS
muon fraction due to the muonic force cannot
be neglected for scalar-muon coupling
$g_{\phi\mu}\gtrsim 10^{-19}$ 
in the long-range force regime, 
where the Compton wavelength far exceeds typical NS radii. 
For example, 
the muon fraction
decreases to one-third of its Standard Model (SM) 
value for the $g_{\phi\mu}= 10^{-18}$ case.

By carefully taking into account the 
effects on the muon fraction from the long-range force, 
we find that the pulsar periastron advance 
provides the most stringent constraints and
is sensitive to the muonic force 
for the mediator mass between
$10^{-18}$ and $\mathcal{O}(10^{-14})$ eV. 
In this parameter space, the muonic force
was previously constrained by 
gravitational wave signal from the NS-NS merger event
detected by the LIGO/Virgo collaboration
\cite{Dror:2019uea,LIGOScientific:2017vwq}.
We find that the periastron advance 
of binary pulsars provides the most stringent 
constraints on the mediator mass range
$\simeq(10^{-17},\,2\times10^{-15})$ eV,
surpassing the GW170817 constraints
by up to one order of magnitude.

The rest of the paper is organized as follows. 
In section \ref{sec:dynamics} we analyze 
the pulsar periastron advance induced by 
ultralight mediators coupled to muons. 
In section \ref{sec:muoninNS} we compute 
the muon fraction, by properly taking into account 
the effects of the long-range Yukawa potential. 
In section \ref{sec:orb-cons}, we compute the constraints
on scalar-muon couplings from the orbital decay measurements of binary
pulsars. 
In section \ref{sec:results} we present the 
pulsar periastron advance constraints and 
the orbital decay constraints on muonic couplings. 
We summarize our findings in section \ref{sec:suma}.  
Some detailed analyses are given in appendices, 
including the neutron
star structure and particle number densities 
in appendix \ref{sec:app-muon}, 
and expressions of the radiation power 
in appendix \ref{sec:app-orb-decay}.

\clearpage

\section{New muonic force and periastron advance}
\label{sec:dynamics}

In this section 
we consider a new muonic force
between the pulsar and its companion NS in binary systems
and study its effects on the periastron advance.

\subsection{New ultralight scalar}

We consider an ultralight scalar $\phi$ 
that couples to muons via the following 
interaction Lagrangian
\begin{equation}\label{eq:lagrangian}
    \mathcal{L}_{\rm int} \supset g_{\phi\mu}\phi \bar{\mu}\mu,
\end{equation}
where $g_{\phi\mu}$ is the dimensionless coupling constant.
We note that our analysis of the periastron
advance for the scalar mediator 
is readily applicable to the vector mediator.

Because 
NS can have a substantial muon fraction \cite{Haensel:2007yy,Pearson:2018tkr}, 
the new muonic ultralight mediator $\phi$ then results in
a long-range potential for muons inside NSs
\begin{equation}
    V(R)=\frac{Q}{4\pi R}e^{-R/\lambda},
\end{equation}
where $Q=g_{\phi\mu}N_\mu$ is the muonic charge of the NS
with $N_\mu$ being the total muon number inside the NS, 
$R$ is the radial distance from the center of the NS, and
$\lambda=1/m_\phi$ is the Compton wavelength of the
light mediator with $m_\phi$ being the mediator mass. 
Thus, in this case, 
muons experience an additional muonic Yukawa potential 
in addition to the gravitational potential.
We compute the muon number of NSs in section 
\ref{sec:muoninNS}. 
We find that, 
{in the absence of any additional muonic interaction}, 
the total muon number is $N_\mu \simeq3\times 10^{55}$
for a typical NS with a mass of $m_{\rm NS}=1.4\, M_\odot$.

For a binary system of a pulsar and an NS, 
the total force between the two orbiting stars 
can then be written as follows
\begin{equation}\label{eq:force}
F(R) = \frac{Gm_1m_2}{R^2}
\left[1-\alpha\,
e^{-\frac{R}{\lambda}}\left(1+\frac{R}{\lambda}\right)\right],
\end{equation}
where $G$ is the gravitational constant, 
$R$ is the distance between the two stars,
$m_1$ and $m_2$ are the masses of the 
pulsar and its companion NS, respectively, 
and
\begin{equation}\label{eq:alpha}
    \alpha\equiv\frac{Q_1 Q_2}{4\pi Gm_1m_2},
\end{equation}
is the dimensionless parameter characterizing 
the relative strength of the muonic force 
with $Q_1$ and $Q_2$ 
being the muonic charges of 
the pulsar and its companion NS, respectively. 
Note that if $\lambda$ is comparable to or larger than
the \zb{semimajor} axis of the binary system, 
the new muonic long-range force in Eq.~\eqref{eq:force} 
is not exponentially suppressed 
and can thus be quite substantial.
For instance,
the Hulse-Taylor binary pulsar, 
PSR B1913+16 \cite{Hulse:1974eb}, 
has a \zb{semimajor} axis of $a\simeq 2\times10^{6}$ km, 
which can be inferred from its orbital period $P_b$
\cite{Weisberg:2016jye} via the Kepler's third law
$a=\left[GMP_b^2/(4\pi^2)\right]^{1/3}$.
Thus, to generate a substantial muonic force between
the Hulse-Taylor pulsar and its companion (NS), the mass 
of the scalar is $\lesssim 10^{-16}$ eV.

\subsection{Periastron advance}

The new muonic force provides a new contribution 
to the periastron advance of the binary systems.   
We denote the position vectors of 
two orbiting stars in the binary system
with $\vec{R}_1$ and $\vec{R}_2$, respectively.
In the center of mass (CM) frame of the binary system, 
the equation of motion is \cite{Poisson_Will_2014}
\begin{equation}
\vec{a}=-\frac{GM}{R^2}\vn+\delta\vec{a},
\end{equation} 
where 
$M=m_1+m_2$ is the total mass of the binary system, 
$\vec R=\vec{R}_1-\vec{R}_2$, 
$\vec{a}\equiv d^2\vec{R}/dt^2$, 
$\vec n = \vec R/R$, and 
$\delta \vec{a}$ is the acceleration 
beyond the Newtonian case.  
Here, $\delta \vec{a}$ can be 
induced both by GR and by new physics:
\begin{equation}
    \delta\vec{a}=\vec{a}_{\rm PN}+\vec{a}_\phi,
\end{equation}
where 
$\vec{a}_{\rm PN}$ is the post-Newtonian (PN) contribution, 
and 
\begin{equation}
\vec{a}_\phi=\frac{GM}{R^2}\alpha\,
    e^{-\frac{R}{\lambda}}\left(1+\frac{R}{\lambda}\right)\vec{n} 
\end{equation}
is the new physics contribution.  
The 1PN contribution is \cite{Poisson_Will_2014,Blanchet:2013haa}
\begin{align}
    \vec{a}_{\rm PN}=&-\frac{GM}{R^2}\bigg\{\Big[(1+3\eta)v^2
    -\frac{3}{2}\eta\dot{R}^2
    \nonumber\\
    &-2(2+\eta)\frac{GM}{R}\Big]\vec{n}
    -\Big[2(2-\eta)\dot{R}\Big]\vec{v}\bigg\},  
\end{align}
where $\eta=m_1m_2/M^2$, 
$\vec{v}=d\vec{R}/dt$ 
is the relative velocity, and 
$v=|\vec{v}|$.

\begin{table*}[t]
    \centering
    \renewcommand{\arraystretch}{1.3}
    \begin{tabular}{|c|c|c|c|c|c|c|}
    \hline
    PSR  & $\dot{\omega}_{\rm 1PN}$ (deg/yr) & $\dot{\omega}_{\rm 2PN}$ (deg/yr) & $\dot{\omega}_{\rm 3PN}$ (deg/yr) & 
    $\dot{\omega}_{\rm data}$ (deg/yr) & $m_1$ ($M_\odot$) & $m_2$ ($M_\odot$)
    \\ \hline
    J0737-3039A \cite{Kramer:2021jcw} & $16.899139$ & $0.000440$
    &$1.8\times10^{-8}$ & $16.899323(13)$ & $1.338$ & $1.249$
    \\ \hline
    B1916+13 \cite{Weisberg:2016jye} & $4.226226$ & $0.000098$
    & $3.2\times10^{-9}$ & $4.226585(4)$ & $1.438$ & $1.390$
    \\ \hline
    J1906+0746 \cite{vanLeeuwen:2014sca} & $7.5836$ & $0.0001$
    & $4.3\times10^{-9}$ &$7.5841(5)$ & $1.291$ & $1.322$
    \\ \hline
    J1756-2251 \cite{Ferdman:2014rna} & $2.58306$ & $0.00003$
    & $6.5\times10^{-10}$ & $2.58240(4)$ & $1.341$ & $1.230$
    \\ \hline
    B2127+11C \cite{Jacoby:2006dy} & $4.4635$ & $0.0001$
    & $4.2\times10^{-9}$ & $4.4644(1)$ & $1.358$ & $1.354$
    \\ \hline
    B1534+12 \cite{Fonseca:2014qla} & $1.7560312$ & $0.0000199$
    &$3.6\times10^{-10}$ & $1.7557950(19)$ & $1.333$ &$1.346$ 
    \\ \hline
\end{tabular}
\caption{The observed periastron advance rate, 
$\dot{\omega}_{\rm data}$, 
and the 
PN contributions for six different pulsar systems.  
The 1PN, 2PN, and 3PN contributions are computed via Eq.~\eqref{eq:omega1PN}, Eq.~\eqref{eq:2pnomega}, 
and equation (539b) of Ref.~\cite{Blanchet:2013haa}, 
respectively. 
The numbers in the parentheses indicate 
the $1\sigma$ uncertainties of $\dot{\omega}_{\rm data}$. 
The masses 
(rounded to four significant figures) 
shown in the last two columns
are from the references in the first column, 
and are used to compute the PN contributions
in this table. 
}
\label{tab:pulsars-PN}
\end{table*}

\begin{table*}[htbp]
    \centering
    \renewcommand{\arraystretch}{1.3}
    \begin{tabular}{|c|c|c|c|c|c|}
    \hline
    PSR  & $\gamma$ (ms) & $s$ & $\dot{P}^{\rm int}_b$ $(10^{-12})$ & 
    $m_1$ ($M_\odot$) & $m_2$ ($M_\odot$)
    \\ \hline
    J0737-3039A \cite{Kramer:2021jcw} & - 
    & $0.999936(9,10)$ &$-1.247782(79)$ 
    & $(1.33801, 1.33830)$ & $(1.24878, 1.24894)$ 
    \\ \hline
    B1916+13 \cite{Weisberg:2016jye} & $4.307(4)$
    & - & $-2.398(0.004)$
    & $(1.430, 1.442)$ & $(1.387, 1.393)$
    \\ \hline
    J1906+0746 \cite{vanLeeuwen:2014sca} & $0.470(5)$ & -
    & $0.55(3)$ &$(1.14, 1.41)$ & $(1.26, 1.37)$ 
    \\ \hline
    J1756-2251 \cite{Ferdman:2014rna} & $1.148(9)$ & -
    & $-0.234(6,9)$ & $(1.33, 1.51)$ & $(1.23, 1.30)$ 
    \\ \hline
    B2127+11C \cite{Jacoby:2006dy} & $4.78(4)$ & -
    & $-3.95(13)$ & $(1.28, 1.46)$ & $(1.32, 1.40)$ 
    \\ \hline
    B1534+12 \cite{Fonseca:2014qla} & $2.0708(5)$ & $0.9772(16)$
    &-& $(1.32,1.40)$ & $(1.34, 1.37)$ 
    \\ \hline
\end{tabular}
\caption{The masses for the pulsar $(m_1)$ and
its companion NS $(m_2)$, 
along with the PK parameters used to derive the masses, 
for six binary systems. 
The measured values of
$x$ are $1.415028603(92)$ s for
PSR J0737-3039A \cite{Kramer:2021jcw}, and
$3.7294636(6)$ for PSR B1534+12 \cite{Fonseca:2014qla}.
For the PK parameters, numbers in parentheses
denote the $1\sigma$ uncertainties; 
for the masses, numbers in parentheses
denote the $2\sigma$ uncertainties.}
\label{tab:pulsars}
\end{table*}

In the presence of $\delta\vec{a}$, the 
longitude of the periastron, 
denoted by $\omega$, 
advances as the stars orbit each other, 
known as the periastron advance. 
After the pulsar completes one orbit 
with the period $P_b$, 
the change in the angle $\omega$ is 
\begin{equation}\label{eq:totpa}
    \Delta\omega=\Delta\omega_\phi+\Delta\omega_{\rm PN},
\end{equation}
where \cite{Poisson_Will_2014,Will:2018bme}
\begin{equation}\label{eq:omegaV}
    \begin{split}
        \Delta\omega_\phi=&\int_0^{2\pi}df \left(\frac{d\omega}{df}\right)_\phi,\\
        \left(\frac{d\omega}{df}\right)_\phi=&-\frac{a^2(1-e^2)^2}{GMe}
    \frac{\cos f}{(1+e\cos f)^2}(\vec{a}_\phi\dotproduct\vec{n}),
    \end{split}
\end{equation}
where $f$ is the true anomaly for the eccentric
orbit, $a=[G M (P_b/2 \pi) ^2]^{1/3}$ is the
\zb{semimajor} axis of the orbit,
and $e$ is the orbital eccentricity. 
The 1PN contribution to the 
periastron advance is 
\cite{Poisson_Will_2014}
\begin{equation}\label{eq:omega1PN}
    \Delta\omega_{\rm 1PN}=\frac{6\pi GM}{a(1-e^2)}.
\end{equation}
The 2PN contribution to the 
periastron advance is 
\cite{Damour:1988mr,Kramer:2021jcw}
\begin{equation}\label{eq:2pnomega}
\begin{split}
    \Delta\omega_{\rm 2PN}=&\frac{6\pi}{1-e^2}f_O\beta_O^4,\\
    \beta_O=&\left(\frac{2\pi GM}{P_b}\right)^{1/3},\\
    f_O=&\frac{1}{1-e^2}\left[\frac{39}{4} \left(\frac{m_1}{M}\right)^2+\frac{15 m_1 m_2}{M^2}+\frac{27}{4} \left(\frac{m_2}{M}\right)^2\right]\\
    &-\left[\frac{13}{4} \left(\frac{m_1}{M}\right)^2+\frac{13 m_1 m_2}{3 M^2}+\frac{1}{4} \left(\frac{m_2}{M}\right)^2\right].
\end{split}
\end{equation}
We neglect higher order PN contributions, as they 
are much smaller than the precision of the 
binary pulsar systems considered in our analysis, 
as shown in 
Table~\ref{tab:pulsars-PN}.

The periastron advance rate is 
\begin{equation}\label{eq:omegadott}
\dot{\omega} 
=\frac{\Delta{\omega}}{P_b}
=\frac{\Delta{\omega}_{\rm PN}(m_1,m_2)+\Delta{\omega}_\phi(m_1,m_2,\alpha)}{P_b}, 
\end{equation}
where 
$\Delta{\omega}_{\rm PN}$ is 
given in Eqs.~(\ref{eq:omega1PN}-\ref{eq:2pnomega}), and 
$\Delta{\omega}_\phi$ is 
given in Eq.~\eqref{eq:omegaV}.

\subsection{Binary masses}
\label{sec:bi-masses}

To compute the periastron advance rate, one has to 
determine the masses of the pulsar and its
companion. 
In the DD timing model 
\cite{AIHPA_1986__44_3_263_0}
used for the data analysis of the binary pulsars
\cite{Weisberg:2016jye, Kramer:2021jcw, vanLeeuwen:2014sca,Ferdman:2014rna,Jacoby:2006dy,Fonseca:2014qla}, 
the masses of the binary stars are derived from 
the measured post-Keplerian (PK) parameters 
\cite{Damour:1991rd}. 
\footnote{For example,
the masses of PSR B1916+13 can be determined by
$\{\dot{\omega},\dot{P}^{\rm int}_b,\gamma,s,r\}$,
as shown in figure 4 of \cite{Weisberg:2016jye}, 
where $r$ is the range of the Shapiro delay
\cite{Damour:1991rd}.} 
In our analysis we consider the following four 
PK parameters for the mass determination of the 
binary pulsar systems 
\footnote{A detailed description of various PK parameters 
can be found in Ref.~\cite{Damour:1991rd}.}: 
\begin{itemize}
\item 
The periastron advance rate $\dot{\omega}$.

\item 
The amplitude of the gravitational 
redshift - time dilation term $\gamma$
\cite{Weisberg:2016jye}: 
\begin{equation}\label{eq:gamma}
    \gamma=e\left(\frac{P_b}{2\pi}\right)^{1/3}T^{2/3}_\odot M^{-4/3}m_2(m_1+2m_2),
\end{equation}
where $T_\odot\equiv GM_\odot$ \cite{Weisberg:2016jye}. 

\item 
The shape of the Shapiro delay $s$
\cite{Kramer:2021jcw}: 
\begin{equation}\label{eq:s}
    s=\sin i=\frac{2\pi xM}{P_b\beta_Om_2}\left[1+\left(3-\frac{m_1m_2}{3M^2}\right)\beta_O^2\right],
\end{equation}
where $\beta_O=\left(2\pi GM/P_b\right)^{1/3}$, and 
$x$ is the projected \zb{semimajor} axis.

\item 
The intrinsic orbital decay rate
$\dot{P}^{\rm int}_b$ 
\cite{Weisberg:2016jye}: 
\begin{equation}\label{eq:Pbdot}
\begin{split}
    \dot{P}^{\rm int}_b=&-\frac{192\pi}{5}\left(\frac{P_b}{2\pi}\right)^{-5/3}
    \left(1+\frac{73}{24}e^2+\frac{37}{96}e^4\right)\\
    &\times (1-e^2)^{-7/2}
    T_\odot^{5/3}m_2m_1M^{-1/3},
\end{split}
\end{equation}
which accounts for the
quadrupole gravitational radiation. 
Note that $\dot{P}^{\rm int}_b$ is
obtained from the observed orbital decay by
removing external contributions, such as the
difference in the galactic gravitational
accelerations between the barycenter of the
solar system and the binary pulsar system
\cite{Kramer:2021jcw}. 
For PSR J0737-3039,
we also consider the higher-order contributions to $\dot{P}^{\rm int}_b$ 
due to the $3.5$PN contribution in the equations of motion, 
which is 
$-1.75\times10^{-17}$ \cite{Kramer:2021jcw}. 

\end{itemize}

For the six binary pulsars we considered, 
$\dot{\omega}$ is the PK parameter that leads to 
the best-measured masses in GR 
\cite{Kramer:2021jcw, Weisberg:2016jye,vanLeeuwen:2014sca,Ferdman:2014rna,Jacoby:2006dy,Fonseca:2014qla}; 
the masses obtained using $\dot{\omega}$ are shown in
Table \ref{tab:pulsars-PN}. 
In our analysis, however, we do not use $\dot{\omega}$ 
to determine the masses, since we use $\dot{\omega}$ 
to compute constraints on new physics. 
Instead we use $\gamma$, $s$, and $\dot{P}^{\rm int}_b$ 
to determine the masses; 
we select two of the three PK parameters that lead to 
the masses that have the smallest uncertainties. 
The obtained masses and 
the PK parameters used to compute the masses are 
shown in Table \ref{tab:pulsars}. 
We find that the masses in Table \ref{tab:pulsars} 
are consistent with the masses in Table \ref{tab:pulsars-PN}.

\subsection{Binary pulsars}

Table~\ref{tab:pulsars-PN} shows the observed 
periastron advance rate and the 1PN, 2PN and 3PN 
contributions for six different binary pulsar systems. 
The 2PN contribution is necessary for 
some of the binary pulsar systems in our analysis, 
since they have highly precise timing data. 
For example, the periastron advance
rate of the double pulsar system PSR J0737-3039A
has been measured with an uncertainty of
$7.7\times 10^{-7}$ times its measured value. 
Note that for PSR J0737-3039A,
we also incorporate the
spin-orbit coupling effect, 
which is $-4.83^{+0.29}_{-0.35}\times 10^{-4}$ 
deg/yr \cite{Kramer:2021jcw}.  
As shown in Table~\ref{tab:pulsars-PN}, 
the 3PN corrections are smaller than the experimental 
uncertainties and thus can be neglected.

\begin{figure}[htbp]
\begin{centering}
\includegraphics[width=0.35 \textwidth]{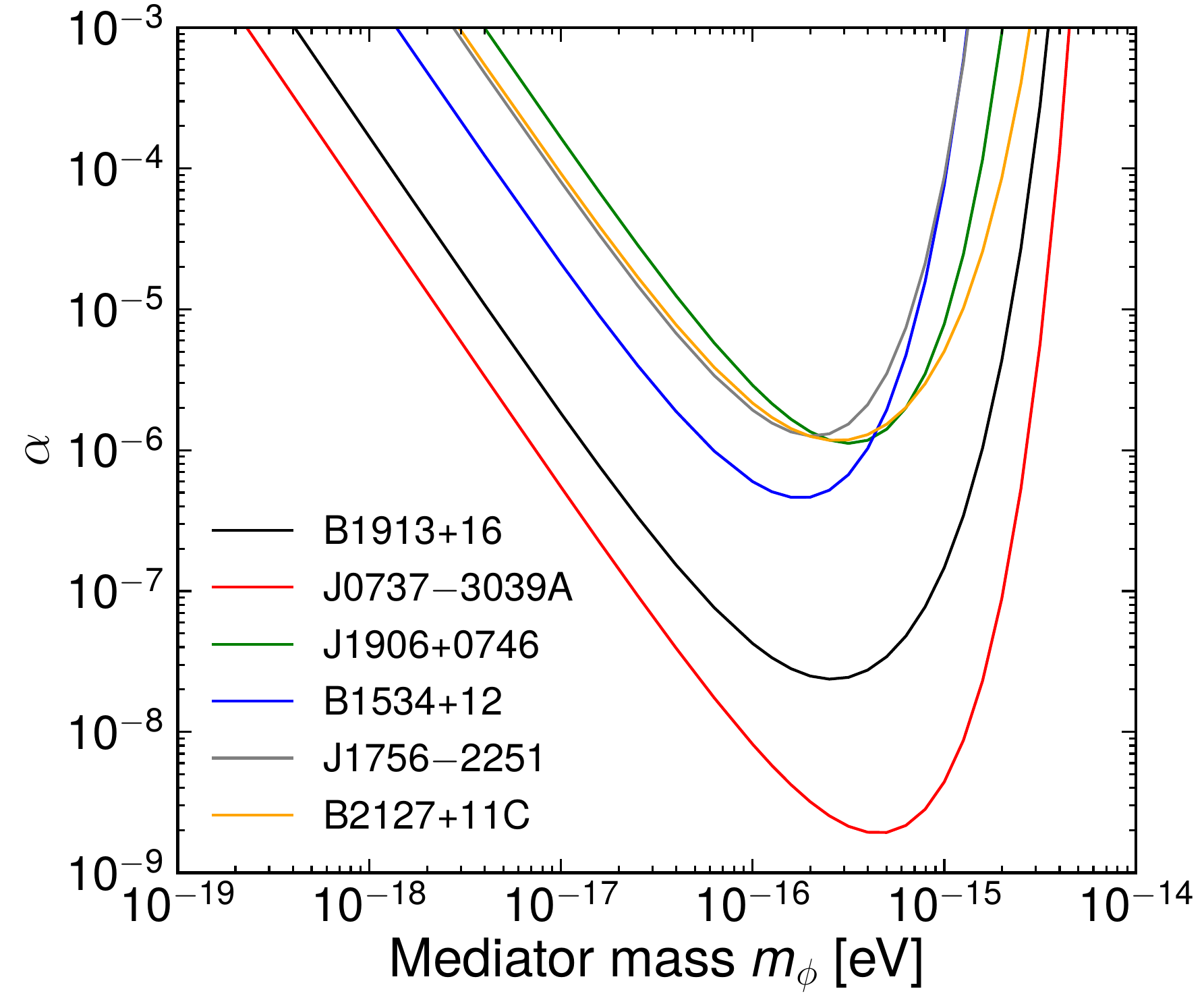}
\caption{The $2\sigma$ bounds on $\alpha$ from
the periastron advance of binary pulsar systems.}
\label{fig:alphaall}
\end{centering}
\end{figure}

To compute the constraints,
we use Eq.~\eqref{eq:omegadott},
where the observed values of $\dot{\omega}$ from
the pulsar-NS binary systems are given in 
Table~\ref{tab:pulsars-PN}, and the 
component masses are given in 
Table \ref{tab:pulsars}. 
Fig.~\ref{fig:alphaall} shows the 
upper bound at the $2\sigma$ level on 
the parameter $\alpha$. 
Among the six binary systems we consider in Table~\ref{tab:pulsars-PN}, 
PSR J0737-3039A has the most accurate timing measurements, 
leading to the most stringent constraints on $\alpha$.

\section{Muon fraction of neutron star}
\label{sec:muoninNS}

In this section we further compute the constraints
on the muonic coupling $g_{\phi\mu}$ 
by using the constraints on $\alpha$ shown in Fig.~\ref{fig:alphaall}.
To do so, we compute
the charge-to-mass ratio $Q/m$ of an NS via
\begin{equation}
\frac{Q}{m}=
\frac{g_{\phi \mu}}{m_n} \frac{N_\mu}{N_n}
\equiv 
\frac{g_{\phi \mu}}{m_n}Y_\mu,  
\end{equation}
where 
$N_\mu$ and $N_n$ are the total number of 
muons and 
nucleons inside the NS, respectively, 
$m_n \simeq 940$ MeV 
is the mass of the neutron, and 
$Y_\mu\equiv N_\mu/N_n$ is the muon fraction. 
In our analysis for the NS, 
we use the 
equation-of-state (EoS) model BSk24 \cite{Pearson:2018tkr}, 
which is the EoS model in the
analysis of the binary merger event GW230529
\cite{LIGOScientific:2024elc}. 
The nucleon number inside an NS is given by   
$N_n = m_{\rm NS}/(e_{\rm eq}+m_n)$, 
where $e_{\rm eq}$ is the energy per nucleon 
without the nucleon mass \cite{Pearson:2018tkr}, 
and we have neglected the mass difference between neutron and proton. 
Note that the pulsars in our analysis have $e_{\rm eq} \ll m_n$;  
for instance, $e_{\rm eq}$ of PSR J0737-3039A 
is in the range of $\sim (62.7,-9.0)$ MeV 
from the center to the outer crust. 
Thus, we further neglect $e_{\rm eq}$ and use 
$N_n \simeq m_{\rm NS}/m_n$ 
to determine the total nucleon number.

We next compute the 
muon fraction $Y_\mu$ inside an NS 
in the presence of an ultralight mediator $\phi$. 
For that purpose, 
one has to account for the effects from the Yukawa potential 
induced by the ultralight mediator on the muon fraction.  
Although such effects have been mentioned in 
Ref.~\cite{Dror:2019uea}, their significance has been 
somewhat overlooked and thus neglected. 
We find that such effects  
are significant in the parameter space of interest 
so that they must be taken into account properly in the calculation.

Neutron stars are primarily composed of 
neutrons, protons, electrons, and muons, 
known as the $npe\mu$ matter 
\cite{Haensel:2007yy}. 
Muons can be produced in the NS core
if the Fermi energy of electrons
exceeds the rest mass of the muon, 
$m_\mu=105.65$ MeV.
Because the Fermi energy of electrons
can be as high as $\sim 122$ MeV 
in the NS core \cite{Haensel:2007yy},
a substantial muon abundance 
is expected in the NS core.

If the $npe\mu$ matter is at equilibrium,
the chemical potential of the muon 
is equal to  
the chemical potential of the electron   
\cite{Haensel:2007yy,Pearson:2018tkr}:
\begin{equation}\label{eq:mubalance}
    \mu_e=\mu_\mu. 
\end{equation}
The chemical potential of the electron 
is given by 
\begin{equation}
\label{eq:mue}
\mu_e(r)=\sqrt{m_e^2+[3\pi^2n_e(r)]^{2/3}},    
\end{equation}
where $n_e(r)$ is the
number density of the electron 
at the radial coordinate $r$ of the NS. 
The chemical potential of the muon  
is given by
\begin{equation}
\label{eq:mumu}
\mu_\mu(r)=\sqrt{m_\mu^2+[3\pi^2n_\mu(r)]^{2/3}} + V_\phi (r),    
\end{equation}
where $n_\mu(r)$ is the muon
number density at the radial coordinate $r$ of the NS, 
and $V_\phi (r)$ is the potential energy  
due to the ultralight mediator $\phi$. 
In the parameter space of interest
we have 
\footnote{Our analysis on $V_\phi$ differs from 
Ref.~\cite{Dror:2019uea} (equation 2 therein) 
by an overall sign.}
\begin{equation}\label{eq:Vphi}
V_\phi (r) = g_{\phi\mu}^2 
\int^\infty_r 
\frac{dr'}{r'^2}\int_0^{r'}dr'' r''^2n_\mu(r''). 
\end{equation}
Here we have approximated the Yukawa potential 
induced by the ultralight mediator $\phi$ with the Coulomb potential. 
This approximation is justified because the Compton wavelength probed by 
the periastron advance measurement is comparable to 
the separation between the two neutron stars in the binary system, 
which is typically much larger than the radius of an NS. 
For example, the separation between the two NSs 
in PSR J0737-3039 is $\sim 10^6$ km, 
whereas the radii of the NSs are $\sim 10$ km.  
Therefore, the Coulomb potential provides a reasonable 
approximation of the 
Yukawa potential when computing the muon distributions 
within the NS.
We provide a detailed derivation of
Eqs.~(\ref{eq:mubalance}-\ref{eq:mumu}), 
and the electron density profile $n_e(r)$ 
in Appendix \ref{sec:app-muon}.

Because $\mu_\mu$ increases with $g_{\phi\mu}^2$, 
the amount of energy required 
to produce muons becomes large 
in the parameter region where the coupling constant $g_{\phi \mu}$ is large. 
This then leads to a suppression on
the muon fraction in the large coupling region.

\begin{figure}[htbp]
\begin{centering}
\includegraphics[width=0.35 \textwidth]{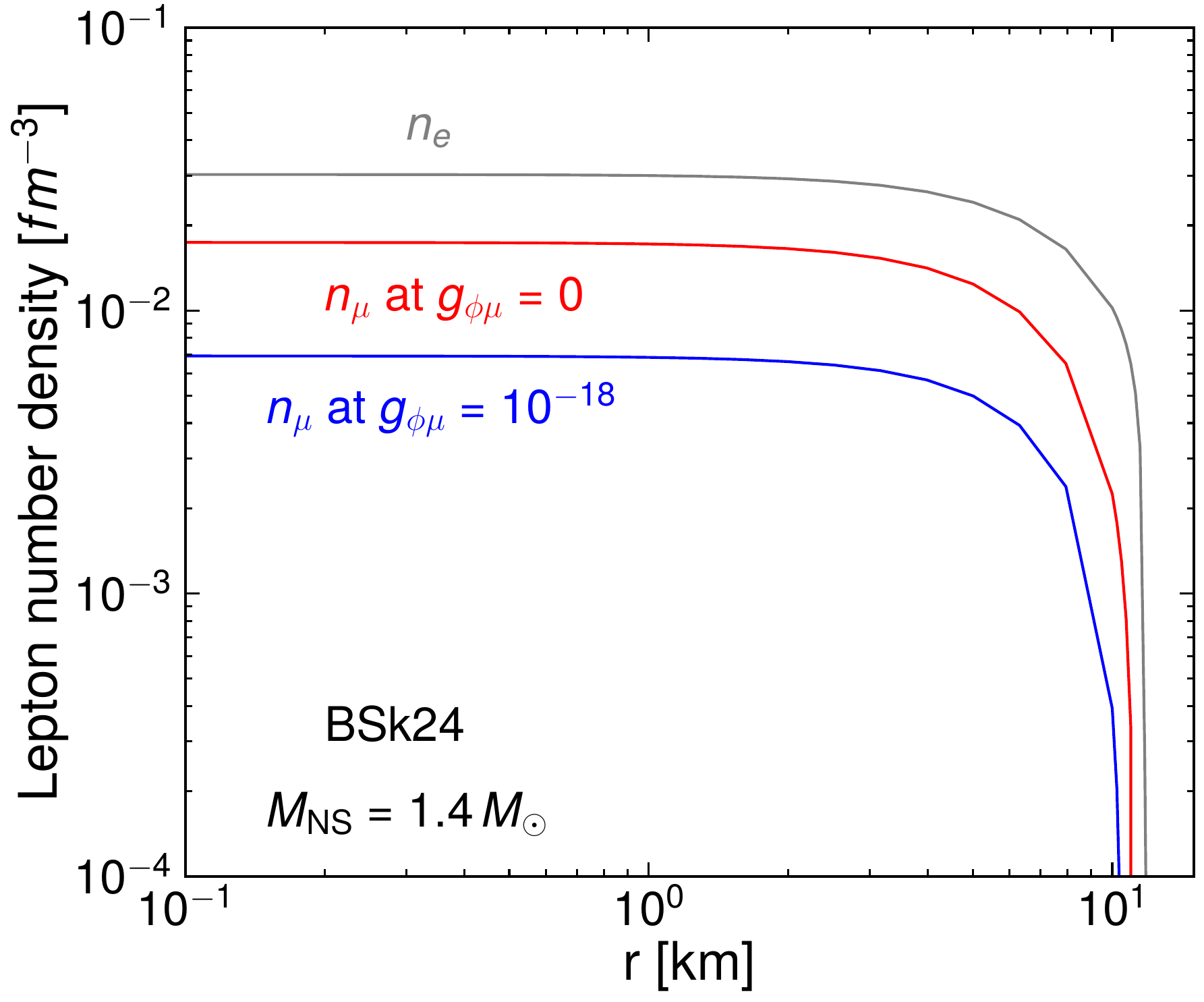}
\caption{Lepton number density as a function of the radial distance $r$ 
for an NS with a mass of $m=1.4\,M_\odot$, 
where the EoS is BSk24 \cite{Pearson:2018tkr}.
The muon number density profiles 
are shown for the $g_{\phi\mu}=10^{-18}$ case (blue) 
and for the $g_{\phi\mu}=0$ case (red). 
Also shown is the electron number density profile (gray).}
\label{fig:muonnr}
\end{centering}
\end{figure}

To determine the profile of the muon number density, 
we employ an iterative
procedure to solve Eq.~(\ref{eq:mumu}) 
such that the muon profile at the $i$-th iteration 
is given by 
\begin{equation}\label{eq:iterationn}
\begin{split}
n_{\mu,i}(r)=&\frac{1}{3\pi^2}
    \Bigg\{\bigg[\mu_e(r) - g_{\phi\mu}^2 
    \int_r^\infty  dr'
    \frac{1}{r'^2}
    \\ &\times 
    \int_0^{r'}dr'' r''^2 
    n_{\mu,i-1}(r'')\bigg]^2-m_\mu^2\Bigg\}^\frac{3}{2},
\end{split}
\end{equation}
where $n_{\mu,i-1}(r)$ is the 
muon profile at the $(i-1)$-th iteration. 
For $n_{\mu,0}(r)$ in the first iteration, 
we use the profile obtained with $g_{\phi\mu}=0$ 
such that 
\begin{equation} 
{m_e^2+[3\pi^2n_e(r)]^{2/3}}
= 
{m_\mu^2+[3\pi^2n_{\mu,0}(r)]^{2/3}}. 
\end{equation} 
The iteration is terminated at
\begin{equation}\label{eq:delmu}
    \delta\mu(r) 
    \equiv 
    \left|
    \frac{\mu_e(r)-\mu_{\mu}(r)}{\mu_e(r)}
    \right| 
    \leq 1\%, 
\end{equation} 
for the $r$ values of interest. 
We consider the range between 
$r=0.1$ km and $r=r_0$ where $r_0$ is given by   
\begin{equation}
    \sqrt{m_e^2+[3\pi^2n_e(r_0)]^{2/3}}=m_\mu.
\end{equation}
We then sample points 
evenly spaced on a logarithmic scale within the range. 
Note that muon production is negligible 
in the region of $r>r_0$, 
where the electron Fermi
energy drops below the muon rest mass. 
Fig.~\ref{fig:muonnr} shows the muon
number density profile for the case where 
$g_{\phi\mu}=10^{-18}$ 
and the NS mass is $1.4\,M_\odot$. 
In this case, we find that 
$\delta\mu(r)<0.2\%$ 
for $r \in (0.1 \text{ km}, r_0)$, 
where $r_0\simeq11.2$ km.
As shown in Fig.~\ref{fig:muonnr}, 
the muon number density in the $g_{\phi\mu}=10^{-18}$ case 
is approximately three times smaller than in the $g_{\phi\mu}=0$ case.

We note that the iterative method in our analysis 
works well in the small coupling regime, 
specifically for $g_{\phi\mu} \lesssim 10^{-18}$.  
However, in the large coupling regime of 
$g_{\phi\mu} \gtrsim 10^{-18}$, 
our iterative method fails to find solutions for $n_\mu$. 
This could be due to the fact that 
as $g_{\phi\mu}$ increases, 
the muonic potential term in Eq.~\eqref{eq:iterationn} 
becomes significant so that it disrupts  
the convergence of the iterative process. 
Consequently, we restrict  
our analysis to $g_{\phi\mu}\leq10^{-18}$ and leave
the investigation of larger couplings for future work.

We then calculate the muon number fraction with the integral
\begin{equation}\label{eq:fracionintegral}
    Y_\mu=\frac{\int_0^{R_{\rm NS}}r^2n_\mu(r)dr }{\int_0^{R_{\rm NS}}r^2n(r)dr},
\end{equation}
where $n(r)$ is the nucleon number density
inside the NS, and $R_{\rm NS}$ is the radius
of the NS. The calculations for $n(r)$ and
$R_{\rm NS}$ are given in Appendix \ref{sec:app-muon}.

\begin{figure}[htbp]
\begin{centering}
\includegraphics[width=0.35 \textwidth]{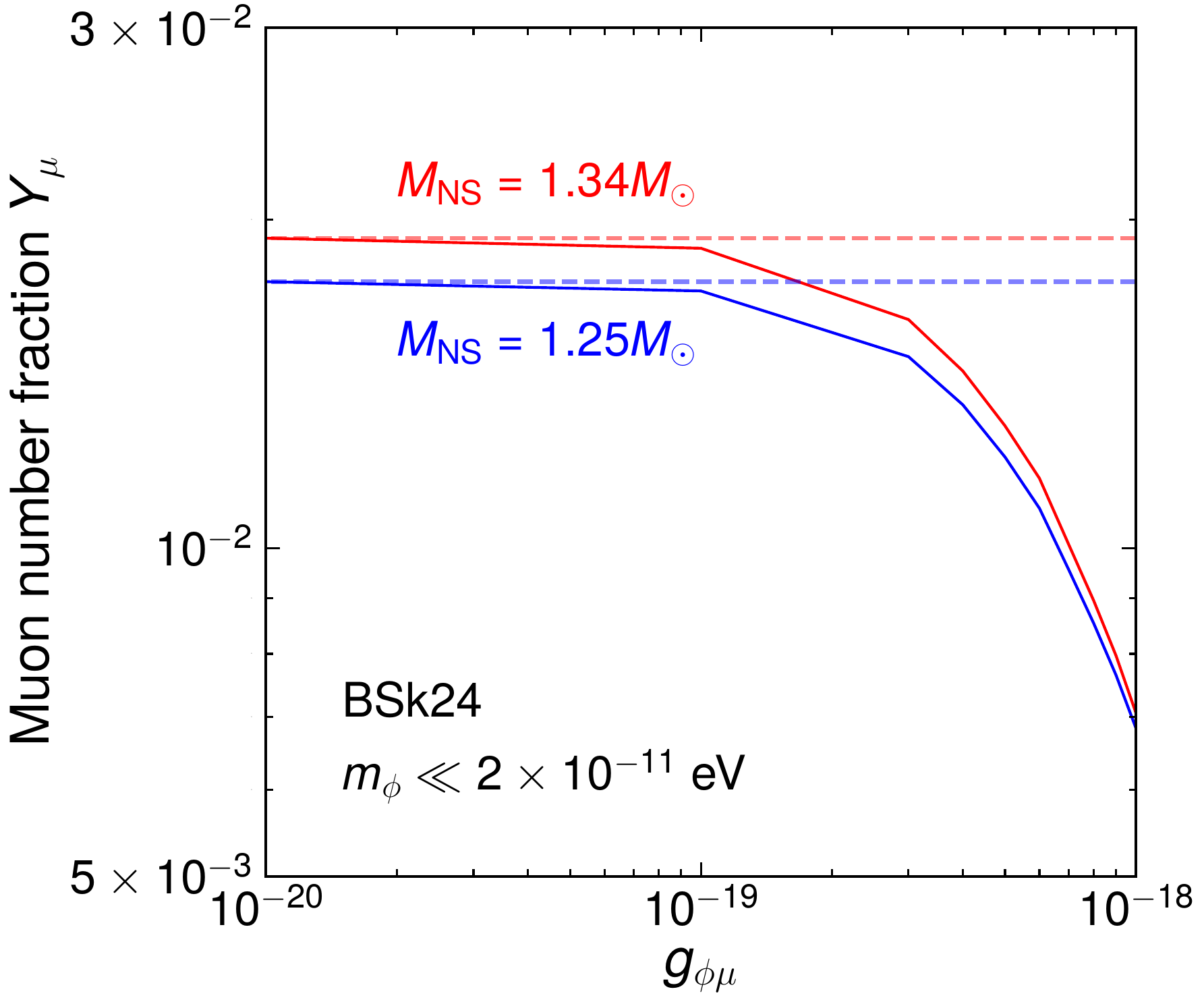}
\caption{The muon number fraction $Y_\mu\equiv N_\mu/N_n$  
as a function of the coupling $g_{\phi\mu}$ 
for NSs with
masses of $1.34\, M_\odot$ (red solid) and $1.25\, M_\odot$ (blue solid), 
in the long-range force regime, 
where the mediator's  
Compton wavelength $\lambda_\phi$ is much larger than the 
NS radius $R_{\rm NS}$; 
for a typical NS radius $R_{\rm NS}\simeq 10$ km \cite{Haensel:2007yy}, 
the condition 
$\lambda_\phi\gg R_{\rm NS}$ corresponds to $m_\phi\ll 2\times10^{-11}$ eV.
The $Y_\mu$ values 
in the $g_{\phi\mu}=0$ cases are shown as dashed lines.
}
\label{fig:muonfraction}
\end{centering}
\end{figure}

Fig.~\ref{fig:muonfraction} shows 
the muon number fraction 
$Y_\mu$ as a function of the coupling $g_{\phi\mu}$, 
in the long-range force regime, where the mediator's  
Compton wavelength far exceeds the typical NS radius.
We find that the new contribution to 
the muon fraction due to the 
muonic potential induced by $\phi$ 
in the long-range force regime 
can no longer be neglected for $g_{\phi\mu}\gtrsim 10^{-19}$.

\section{Pulsar orbital decay constraints}
\label{sec:orb-cons}

In this section we obtain the pulsar orbital decay constraints 
on muonic couplings. 
We attribute the difference between 
the intrinsic orbital decay
{and}  
the predictions of general relativity (GR) to 
dipole radiation from the ultralight mediator. 
To obtain constraints on new physics, 
we use 
\cite{Krause:1994ar}
\begin{equation}\label{eq:Pdotratio}
    \frac{\langle\dot{E}\rangle_{S,V}(\gamma)}{\langle\dot{E}\rangle_{G}}=
    1-\frac{\dot{P}_b^{\rm GR}}{\dot{P}_b^{\rm int}},
\end{equation}
where 
$\dot{P}_b^{\rm GR}$ is the GR-predicted orbital decay,
$\dot{P}_b^{\rm int}$ is the intrinsic orbital decay,
and $\langle\dot{E}\rangle_{S,V}$ and 
$\langle\dot{E}\rangle_{G}$ are the
time-averaged radiation
powers, with $S$, $V$, $G$ denoting
the scalar mediator, vector mediator, 
and gravity, respectively. 
Note that $\langle\dot{E}\rangle_{S,V}$ is 
a function of $\gamma=[(Q/m)_1-(Q/m)_2]^2/(4\pi G)$, 
which is the strength of the dipole radiation.
The expressions for the radiation powers
are given in Appendix \ref{sec:app-orb-decay}. 
Fig.~\ref{fig:OD-bounds} shows 
constraints (at the $2\sigma$ level) 
on $\gamma$ for both the scalar and vector mediators, 
from the binary pulsar systems in Table \ref{tab:pulsars-OD}. 
The strongest bounds come from the latest observation of the double pulsar
system PSR J0737-3039A \cite{Kramer:2021jcw}.

\begin{figure*}[htbp]
\begin{centering}
\includegraphics[width=0.35 \textwidth]{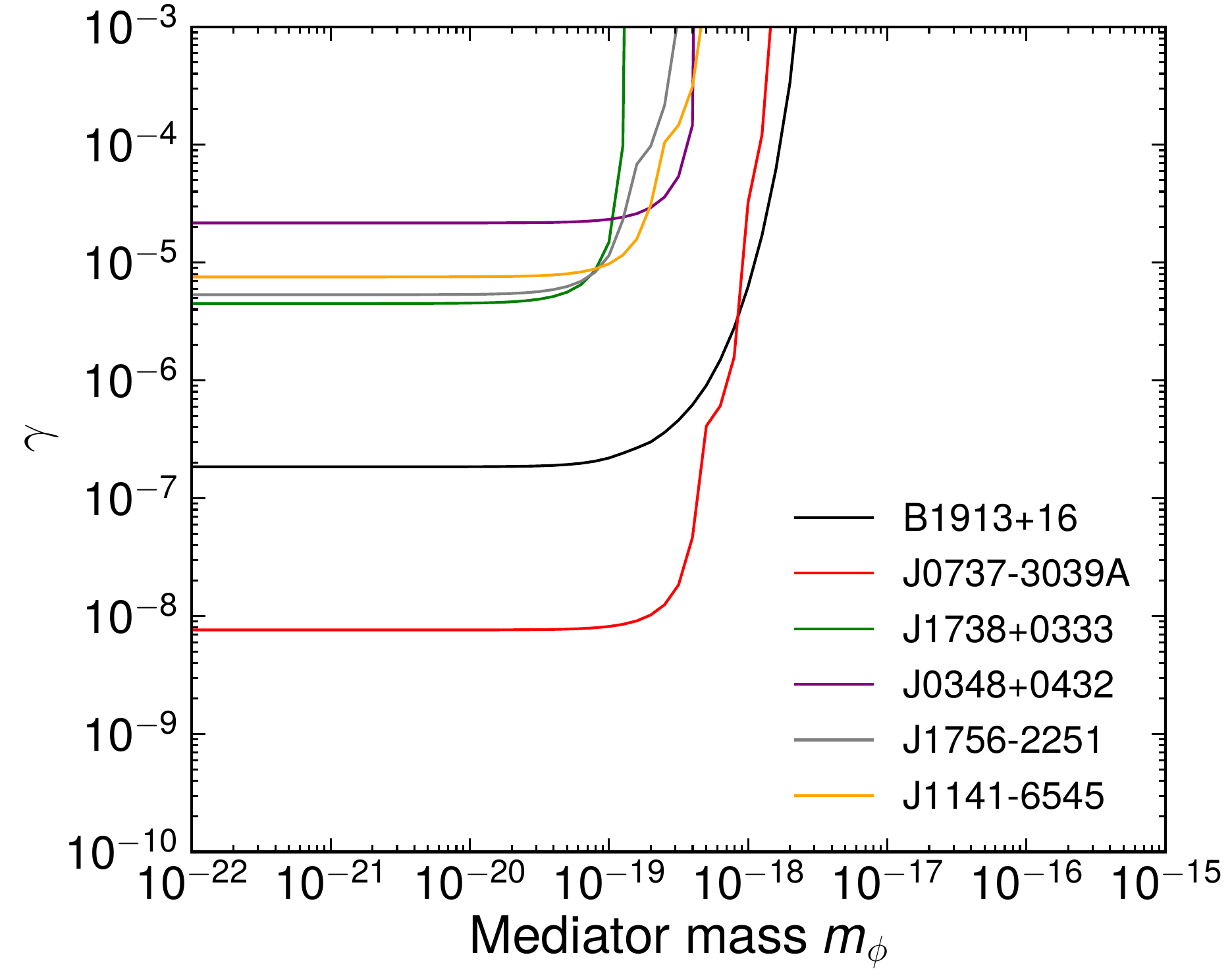}
\hspace{1cm}
\includegraphics[width=0.35 \textwidth]{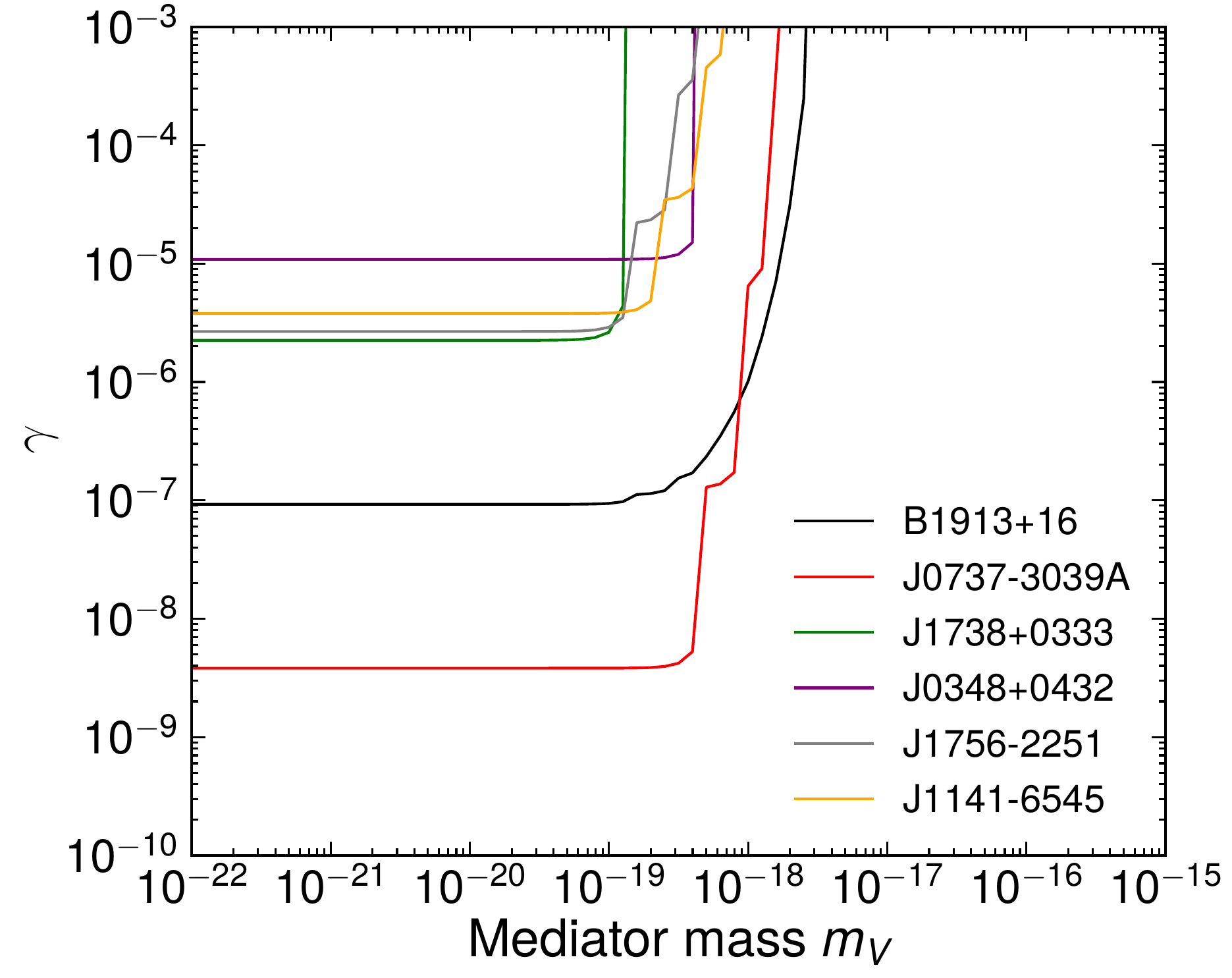}
\caption{The $2\sigma$ upper bounds on $\gamma$ 
from the orbital decay measurements of binary pulsar systems 
for the scalar (left) and vector (right) cases. 
}
\label{fig:OD-bounds}
\end{centering}
\end{figure*}

To compute constraints on the
scalar/vector-muon coupling $g$, we use 
\begin{equation}\label{eq:gammag}
\begin{split}
    \gamma&=\frac{1}{4\pi G}\left[\left(\frac{Q}{m}\right)_1
    -\left(\frac{Q}{m}\right)_2\right]^2\\
    &=\frac{g^2}{4\pi Gm_n^2}\left(Y_{\mu,1}
    -Y_{\mu,2}\right)^2,
\end{split}
\end{equation}
where $Y_{\mu,i}$ are the muon fraction
of the NS or white dwarf (WD) with $i=1,2$
denoting the pulsar and its companion. 
We use Eq.~\eqref{eq:fracionintegral} 
to compute $Y_\mu$ for the NS; 
for white dwarfs, we assume $Y_\mu=0$
\cite{Shapiro:1983du}.

\section{Results}
\label{sec:results}

\begin{figure}[htbp]
\begin{centering}
\includegraphics[width=0.35 \textwidth]{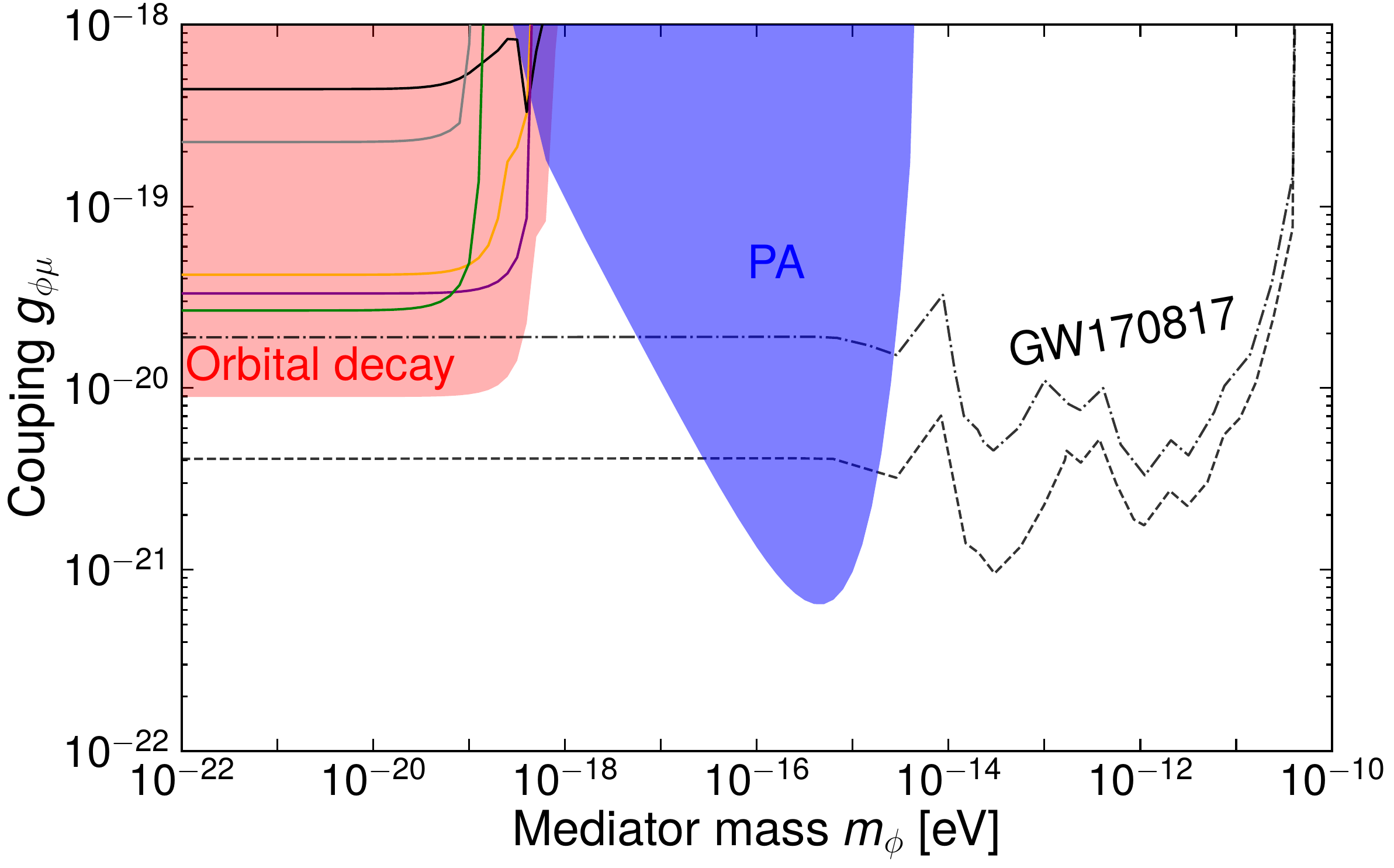}
\caption{The 95\% CL constraints on the
scalar-muon coupling $g_{\phi \mu}$ 
from the periastron advance (PA) of binary pulsar system 
PSR J0737-3039 (blue-shaded), 
and from the orbital decay of binary pulsar systems (red-shaded). 
Also shown are the GW170817 constraints \cite{Dror:2019uea}
computed with four different EoSs; 
the limits with BSk24 lie in between the two 
extreme cases shown as dashed and dash-dotted curves.}
\label{fig:scalar-cons}
\end{centering}
\end{figure}

Fig.~\ref{fig:scalar-cons} shows the  
constraints on the scalar-muon coupling $g_{\phi \mu}$ 
from 
the periastron advance of PSR J0737-3039A 
and 
from 
the orbital decay measurements of 
the binary pulsar systems given in Table~\ref{tab:pulsars-OD}. 
The constraints from NS-WD binaries 
are generally more stringent than those
from NS-NS binaries, except for the  
double pulsar system PSR J0737-3039A. 
That NS-WD binaries provide better constraints 
is primarily due to the vanishing muon fraction in WDs, 
which leads to a typically large 
$\left(Y_{\mu,1}-Y_{\mu,2}\right)^2$ factor in Eq.~\eqref{eq:gammag}. 
The most stringent constraint  
from the double pulsar system PSR J0737-3039A,
is primarily due to 
the high precision measurements of the orbital decay.

We compare our results with 
the GW170817 constraints given in Ref.~\cite{Dror:2019uea}, 
which are analyzed for the $L_\mu-L_\tau$ gauged boson. 
We obtain the GW170817 limits on $g_{\phi\mu}$ 
(due to dipole radiation)
by rescaling the ones given in Ref.~\cite{Dror:2019uea} 
by a factor of $\sqrt{2}$, 
since the radiation power of the scalar mediator 
is a factor of $2$ smaller than 
that of the vector mediator \cite{Krause:1994ar}, 
leading to a weaker constraint on $\gamma$ by a factor of 2. 
The GW170817 constraints depend on the muon abundance, 
which in turn depends on the EoS. 
Ref.~\cite{Dror:2019uea} analyzed the 
GW170817 limits using four different EoS from 
Ref.~\cite{Pearson:2018tkr}, including 
BSk22, BSk24, BSk25, and BSk26. 
The envelope of these four cases is shown 
in Fig.~\ref{fig:scalar-cons}. 
Note that 
the BSk22 (BSk26) EoS predicts 
the highest (lowest) muon abundance \cite{Dror:2019uea}, 
and the muon abundance predicted by the BSk24 EoS 
lies in between.

As shown in Fig.~\ref{fig:scalar-cons}, 
the pulsar orbital decay constraints
are compatible with the GW170817 constraints
\cite{Dror:2019uea} for
$m_\phi\lesssim 10^{-19}$ eV. 
The constraints from the pulsar periastron
advance are the strongest  
for the mediator mass range of 
$m_\phi\sim(10^{-17},2\times10^{-15})$ eV.

\section{Summary}
\label{sec:suma}

In this paper we compute pulsar periastron advance constraints on
ultralight scalar mediators that couple to muons. 
We consider six pulsar binary systems with precise timing 
measurements, and find that 
the PSR J0737-3039A provides the strongest constraint, 
owing to its highly precise timing measurements. 
The periastron advance provides the most stringent constraints 
in the mass range of
$m_\phi\in(10^{-17},2\times10^{-15})$ eV, 
with the best limit on the scalar muon coupling 
$g_{\phi\mu}<\mathcal{O}(10^{-21})$ at 
$m_\phi\simeq 4\times10^{-16}$ eV; 
this surpasses the GW170817 constraints by
about an order of magnitude,
as shown in Fig.~\ref{fig:scalar-cons}.

The pulsar periastron advance constraints 
on muonic forces 
are strongly dependent on the muon composition 
of the neutron stars. 
We computed the muon fraction of the NSs 
by using the phenomenologically superior
BSk24 EoS \cite{Pearson:2018tkr,Pearson:2020bxz}, which
is consistent with both nuclear physics
and astrophysical constraints,
and is used for analyzing the recently
observed binary compact objects merger event GW230529
\cite{LIGOScientific:2024elc}.
Moreover, we consider the effects of
the long-range muonic force,
which are important
when the coupling $g_{\phi\mu}$ is
larger than $10^{-19}$.
The effects of the long-range muonic force become so significant that 
the muon fraction is reduced to one-third
of its value in SM when $g_{\phi\mu} \sim 10^{-18}$. 
We thus conclude that previous studies 
that neglect the effects of the long-range muonic force 
overestimate the capability of neutron stars to probe 
ultralight muonic mediators with large couplings.

In our analysis we also study  
the constraints on the radiation
strength from orbital decay measurements
of pulsar-NS and pulsar-WD binaries,
for both scalar and vector mediator cases. 
The latest measurement of PSR J0737-3039A provides the best orbital
decay constraints. 
After taking into account the suppression
effects on the muon fraction due to the long-range force, 
we find that 
the orbital decay constraints on the scalar coupling
$g_{\phi\mu}$ exceed the GW170817 constraints that are calculated
under a pessimistic muon fraction assumption \cite{Dror:2019uea}
for $m_\phi\lesssim10^{-19}$ 
{eV}.

\begin{acknowledgments}
We thank Evan McDonough,
Nicol\'{a}s Yunes, Caroline Owen,
and Yiqi Xie for discussions and
correspondence. 
The work is supported in part by the 
National Natural Science Foundation of China 
under Grant No.\ 12275128.

\end{acknowledgments}

\appendix

\section{$npe\mu$ inside neutron star}
\label{sec:app-muon}

In this section we present some detailed 
calculations of NSs, including 
the nucleon number density $n(r)$,
the electron number density $n_e(r)$, 
the radius of the NS $R_{\rm NS}$, 
and the derivation of 
Eqs.~(\ref{eq:mubalance}-\ref{eq:mumu}).

\subsection{Number densities and NS radius}

The detailed calculation of the 
nucleon number density profile $n(r)$ for the BSk24 EoS 
can be found in Appendix A of Ref.~\cite{Liu:2024qzp}. 
Below we provide a brief summary of the analysis.

We first obtain the mass-energy density $\rho(r)$ 
by solving the
Tolman-Oppenheimer-Volkoff (TOV) equation 
\cite{Tolman:1939jz,Oppenheimer:1939ne}, 
together with the EoS BSk24 
\cite{Pearson:2018tkr}. 
This process also yields the mass-radius
relation of the NSs, from which we determine
the radius for a given NS mass. 
We then determine the nucleon number density $n(r)$ 
via \cite{Pearson:2018tkr}
\begin{equation}\label{eq:rho-n}
    \rho(r) =n(r)(e_{\text{eq}}+m_n), 
\end{equation}
where $m_n$ is the neutron mass,
and $e_{\text{eq}}$ is the energy per nucleon
(without the rest mass term).  
We obtain the electron number density 
$n_e(r)$ by using the fitting function
$Y_e \equiv n_e(r)/n(r)$, which is given
by Eq.~(C17) of Ref.~\cite{Pearson:2018tkr}. 
The electron number density $n_e(r)$ is shown
in Fig.~\ref{fig:muonnr}.

\subsection{Chemical potentials at equilibrium}

The equilibrium of the $npe\mu$ matter is
maintained by the following 
weak interaction processes: 
\cite{Haensel:2007yy}
\begin{align}
    \label{eq:scattring1} &n\rightarrow p+e+\bar{\nu}_e,\\
    \label{eq:scattring2} &n\rightarrow p+\mu+\bar{\nu}_\mu,\\
    \label{eq:scattring3} &p+e\rightarrow n+\nu_e,\\
    \label{eq:scattring4} &p+\mu\rightarrow n+\nu_\mu.
\end{align}
At equilibrium, the chemical potentials on both
sides of these interactions are equal, 
leading to
\begin{align}
    \label{eq:scattringp1} &\mu_n=\mu_p+\mu_e,\\
    \label{eq:scattringp2} &\mu_e=\mu_\mu,
\end{align}
where we have neglected neutrino chemical potentials 
because the NS core is assumed to be transparent to neutrinos.

\subsection{Lepton chemical potential inside NS}

The Fermi distribution of the lepton is given by
\cite{kittel1980thermal}
\begin{align}
    n_F(E)=\frac{1}{e^{(E-\mu_\ell)/T}+1},
\end{align}
where $\mu_\ell$ is the lepton chemical potential.

In the presence of a muonic potential,
the energy per lepton inside the NS is 
\begin{equation}\label{eq:energypoential}
    E=\sqrt{m_\mu^2+p^2}+g_{\phi\mu}V,
\end{equation}
where $p$ is the momentum of the lepton, and
$V$ is the muonic potential. 
For electrons, $V=0$.

For a cold NS such that $T\rightarrow 0$, 
the highest energy level with non-vanishing
$n_F$ is $E_{\rm max}=\mu_\ell$. In this case,
the number density of the lepton is given by
\begin{align}
    n&=2\int \frac{d^3p}{(2\pi)^3}\,\frac{1}{e^{(E-\mu_\ell)/T}+1}\nonumber\\
    &=\frac{1}{\pi^2}\int_0^{\sqrt{
    (\mu_\ell-g_{\phi\mu}V)^2-m_\ell^2}}
    dp\, p^2\nonumber\\
    \label{eq:ndes}&=\frac{1}{3\pi^2}[(\mu_\ell-g_{\phi\mu}V)^2-m_\ell^2]^{3/2}.
\end{align}
Inverting Eq.~\eqref{eq:ndes} we find
\begin{equation}\label{eq:chemicalNP}
\begin{split}
    \mu_\ell(r)=\sqrt{m_\ell^2+[3\pi^2n_\ell(r)]^{2/3}}+g_{\phi\mu}V(r).
\end{split}
\end{equation}
Because the NS radius ($\sim 10$ km) is far
smaller than both the \zb{semimajor} axis ($a\gtrsim 10^6$ km) of
the binary systems in our analysis and the
mediator wavelength we are interested in, 
we can approximate
the Yukawa potential with the Coulomb potential 
\begin{equation}
    V(r)=-\int_\infty^rdr'\frac{Q(r')}{4\pi r'^2},
\end{equation}
where $Q(r')$ is the charge carried by all the muons
inside $r'$:
\begin{equation}\label{eq:chargeinr}
    Q(r')=4\pi g_{\phi\mu} \int_0^{r'}dr'' r''^2n_\mu(r'').
\end{equation}

\begin{table*}[t]
    \centering
    \renewcommand{\arraystretch}{1.3}
    \begin{tabular}{|c|c|c|c|c|c|}
    \hline
    PSR  & $\dot{P}_b^{\rm int}/\dot{P}_b^{\rm GR}$ 
    & $P_b$ (day) & $e$ & $m_1$ ($M_\odot$) & $m_2$ ($M_\odot$)
    \\ \hline
    B1916+13 \cite{Weisberg:2016jye} & $0.9983 \pm 0.0016$
    & $0.323$ & $0.617$ & $1.438$ & $1.39$
    \\ \hline
    J0737-3039A \cite{Kramer:2021jcw} & $0.999963\pm 0.000063$
    & $0.102$ & $0.0878$ & $1.338$ & $1.249$
    \\ \hline
    *J1738+0333 \cite{Freire:2012mg} & $0.94\pm 0.13$
    & $0.355$ & $\simeq 3.4\times10^{-7}$ & $1.46$ & $0.181$
    \\ \hline
    *J0348+0432 \cite{Antoniadis:2013pzd} & $1.05\pm 0.18$
    & $0.102$ & $\simeq 2\times10^{-6}$ & $2.01$ & $0.172$
    \\ \hline
    J1756-2251 \cite{Ferdman:2014rna} & $1.08\pm 0.03$
    & $0.320$ &   $0.181$ & $1.341$ & $1.230$
    \\ \hline
    *J1141-6545 \cite{Bhat:2008ck} & $1.04\pm 0.06$
    & $0.198$ &    $0.172$ & $1.27$ & $1.02$
    \\ \hline
\end{tabular}
\caption{Binary pulsar systems used for 
the orbital decay constraints. 
The NS-WD binaries
are marked by the ``*'' symbol.}
\label{tab:pulsars-OD}
\end{table*}

\section{Pulsar orbital decay}
\label{sec:app-orb-decay}

Ultralight mediator can be emitted from
the NSs via dipole radiation \cite{Krause:1994ar}, 
providing a new energy
loss channel beyond the gravitational radiation for binary pulsar systems.

The ratio of the radiation power from new physics to 
that in GR is  
\cite{Krause:1994ar} 
\footnote{Our formulas differ by a factor
of $1/4\pi$ from those 
given in Ref.~\cite{Krause:1994ar}, 
which uses the Gaussian units.}
\begin{equation}\label{eq:Edotratio}
    \frac{\langle\dot{E}\rangle_{S,V}}{\langle\dot{E}\rangle_{G}}=\left\{
    \begin{split}
        &\frac{5}{96}\gamma
        \left(\frac{P_b}{2\pi GM}\right)^{2/3}
        \frac{g_S(m_\phi,e)}{g_G(e)}\,\,\,\,\,\text{scalar}\\
        &\frac{5}{48}\gamma
        \left(\frac{P_b}{2\pi GM}\right)^{2/3}
        \frac{g_V(m_V,e)}{g_G(e)}\,\,\,\,\text{vector}
    \end{split}\right. ,
\end{equation}
where $g_{S,V}$ are given by
\begin{align}
    g_V(m_V,e)=&\sum_{n>n_0}2n^2
    \left[\mathcal{J}'^2_n(ne)+\left(\frac{1-e^2}{e^2}\right)\mathcal{J}^2_n(ne)\right] \nonumber \\ 
    &\times \sqrt{1-\left(\frac{n_0}{n}\right)^2}
    \left[1+\frac{1}{2}\left(\frac{n_0}{n}\right)^2\right],\\
    g_S(m_\phi,e)=&\sum_{n>n_0}2n^2
    \left[\mathcal{J}'^2_n(ne)+\left(\frac{1-e^2}{e^2}\right)\mathcal{J}^2_n(ne)\right] \nonumber \\
    &\times \left[1-\left(\frac{n_0}{n}\right)^2\right]^{3/2},
\end{align}
with $\mathcal{J}_n$ being the
Bessel function of n-th order, 
$\mathcal{J}'_n(x)\equiv d\mathcal{J}_n(x)/dx$, and
$n_0\equiv P_b m_{\phi,V}/(2\pi)$.
The $g_G$ factor is given by\footnote{As shown in 
equation
(547) of
Ref.~\cite{Blanchet:2013haa},
the $0$PN contribution to GW radiation
includes a factor of $x^5$, where $x=(P_b/2\pi GMK)^{-2/3}$ 
with $K=1+k=1+\Delta\omega/2\pi$ being
the number of
orbits the pulsar completes in one period.
Expanding $x^5$ in terms of $k$ results in the
first term in the expression for $g_{G,1\rm PN}$,
which is of the $1$PN order, 
as given in Eq.~\eqref{eq:g-3.5}.}
\begin{align}
    g_G & = g_{G,0\rm PN}+g_{G,1\rm PN}, \\
    \label{eq:g-2.5} g_{G,0\rm PN}&=\frac{1+\frac{73}{24}e^2+\frac{37}{96}e^4}{(1-e^2)^{7/2}},\\
    \label{eq:g-3.5}g_{G, 1\rm PN}&=
    \frac{10k}{3}g_{G,0\rm PN}+
    \left(\frac{P_b}{2\pi GM}\right)^{-2/3}\frac{1}{(1-e^2)^{9/2}}\nonumber \\
    &\times
    \bigg[
    -\frac{1247}{336}-\frac{35}{12}\eta 
    +e^2\left(\frac{10475}{672}-\frac{1081}{36}\eta\right)\nonumber \\
    &+e^4\left(\frac{10043}{384}-\frac{311}{12}\eta\right)
    +e^6\left(\frac{2179}{1792}-\frac{851}{576}\eta\right)
    \bigg],
\intertext{where $g_{G,0\rm PN}$ and $g_{G,1\rm PN}$
account for the $0$PN and $1$PN contributions
to the gravitational radiation \cite{Blanchet:2013haa},
$k=3(P_b/2\pi GM)^{-2/3}(1-e^2)^{-1}$, and 
$\eta=m_1m_2/(m_1+m_2)^2$.}\nonumber
\end{align}

Table \ref{tab:pulsars-OD} shows the
binary pulsar systems used to calculate
the orbital decay constraints.
For each binary system, we provide
the ratio of its intrinsic orbital decay
to that predicted by GR, 
along with other relevant parameters.
Note that the orbital
decay ratio of PSR J0737-3039A, as
given by Ref.~\cite{Kramer:2021jcw},
accounts for higher-order effects not
included in other binary pulsar systems.
These include
(i) the $1$PN contribution to the GW 
radiation in computing $\dot{P}_b^{\rm GR}$,
and (ii) the mass loss effect 
in calculating $\dot{P}_b^{\rm int}$.\footnote{
The intrinsic orbital decay
after removing the mass loss effect is denoted
as $\dot{P}^{\rm GW}_b$ in Ref.~\cite{Kramer:2021jcw}.
For simplicity, we still denote it as 
$\dot{P}^{\rm int}_b$ here.}
Both effects are smaller than the uncertainty
of $\dot{P}_b^{\rm int}$.

\normalem
\bibliography{ref.bib}{}

\providecommand{\href}[2]{#2}\begingroup\raggedright\begin{thebibliography}{10}

\bibitem{Will:2014kxa}
C.~M.~Will, ``{The Confrontation between General Relativity and Experiment},''
  \href{https://dx.doi.org/10.12942/lrr-2014-4}{Living Rev.\  Rel.\  {\bfseries
  17} (2014) 4} {\ttfamily
  [\href{https://arxiv.org/abs/1403.7377}{arXiv:1403.7377}]}.

\bibitem{Tino:2020nla}
G.~M.~Tino, L.~Cacciapuoti, S.~Capozziello, G.~Lambiase, and F.~Sorrentino,
  ``{Precision Gravity Tests and the Einstein Equivalence Principle},''
  \href{https://dx.doi.org/10.1016/j.ppnp.2020.103772}{Prog.\  Part.\  Nucl.\
  Phys.\  {\bfseries 112} (2020) 103772} {\ttfamily
  [\href{https://arxiv.org/abs/2002.02907}{arXiv:2002.02907}]}.

\bibitem{Adelberger:2006dh}
E.~G.~Adelberger, {\em et al.}, ``{Particle Physics Implications of a Recent
  Test of the Gravitational Inverse Sqaure Law},''
  \href{https://dx.doi.org/10.1103/PhysRevLett.98.131104}{Phys.\  Rev.\  Lett.\
   {\bfseries 98} (2007) 131104} {\ttfamily
  [\href{https://arxiv.org/abs/hep-ph/0611223}{hep-ph/0611223}]}.

\bibitem{Schlamminger:2007ht}
S.~Schlamminger, K.~Y.~Choi, T.~A.~Wagner, J.~H.~Gundlach, and
  E.~G.~Adelberger, ``{Test of the equivalence principle using a rotating
  torsion balance},''
  \href{https://dx.doi.org/10.1103/PhysRevLett.100.041101}{Phys.\  Rev.\
  Lett.\  {\bfseries 100} (2008) 041101} {\ttfamily
  [\href{https://arxiv.org/abs/0712.0607}{arXiv:0712.0607}]}.

\bibitem{Wagner:2012ui}
T.~A.~Wagner, S.~Schlamminger, J.~H.~Gundlach, and E.~G.~Adelberger,
  ``{Torsion-balance tests of the weak equivalence principle},''
  \href{https://dx.doi.org/10.1088/0264-9381/29/18/184002}{Class.\  Quant.\
  Grav.\  {\bfseries 29} (2012) 184002} {\ttfamily
  [\href{https://arxiv.org/abs/1207.2442}{arXiv:1207.2442}]}.

\bibitem{Berge:2017ovy}
J.~Berg\'e, {\em et al.}, ``{MICROSCOPE Mission: First Constraints on the
  Violation of the Weak Equivalence Principle by a Light Scalar Dilaton},''
  \href{https://dx.doi.org/10.1103/PhysRevLett.120.141101}{Phys.\  Rev.\
  Lett.\  {\bfseries 120} (2018) 141101} {\ttfamily
  [\href{https://arxiv.org/abs/1712.00483}{arXiv:1712.00483}]}.

\bibitem{Williams:2004qba}
J.~G.~Williams, S.~G.~Turyshev, and D.~H.~Boggs, ``{Progress in lunar laser
  ranging tests of relativistic gravity},''
  \href{https://dx.doi.org/10.1103/PhysRevLett.93.261101}{Phys.\  Rev.\  Lett.\
   {\bfseries 93} (2004) 261101} {\ttfamily
  [\href{https://arxiv.org/abs/gr-qc/0411113}{gr-qc/0411113}]}.

\bibitem{Turyshev:2006gm}
S.~G.~Turyshev and J.~G.~Williams, ``{Space-based tests of gravity with laser
  ranging},'' \href{https://dx.doi.org/10.1142/S0218271807011838}{Int.\  J.\
  Mod.\  Phys.\  D {\bfseries 16} (2007) 2165--2179} {\ttfamily
  [\href{https://arxiv.org/abs/gr-qc/0611095}{gr-qc/0611095}]}.

\bibitem{Haensel:2007yy}
P.~Haensel, A.~Y.~Potekhin, and D.~G.~Yakovlev,
  \href{https://dx.doi.org/10.1007/978-0-387-47301-7}{{\em {Neutron stars 1:
  Equation of state and structure}}}, vol.~326.
\newblock Springer, New York, USA, 2007.

\bibitem{Dror:2019uea}
J.~A.~Dror, R.~Laha, and T.~Opferkuch, ``{Probing muonic forces with neutron
  star binaries},''
  \href{https://dx.doi.org/10.1103/PhysRevD.102.023005}{Phys.\  Rev.\  D
  {\bfseries 102} (2020) 023005} {\ttfamily
  [\href{https://arxiv.org/abs/1909.12845}{arXiv:1909.12845}]}.

\bibitem{Xu:2020qek}
X.-J.~Xu, ``{The $\nu_{R}$-philic scalar: its loop-induced interactions and
  Yukawa forces in LIGO observations},''
  \href{https://dx.doi.org/10.1007/JHEP09(2020)105}{JHEP {\bfseries 09} (2020)
  105} {\ttfamily [\href{https://arxiv.org/abs/2007.01893}{arXiv:2007.01893}]}.

\bibitem{KumarPoddar:2019ceq}
T.~Kumar~Poddar, S.~Mohanty, and S.~Jana, ``{Vector gauge boson radiation from
  compact binary systems in a gauged $L_\mu-L_\tau$ scenario},''
  \href{https://dx.doi.org/10.1103/PhysRevD.100.123023}{Phys.\  Rev.\  D
  {\bfseries 100} (2019) 123023} {\ttfamily
  [\href{https://arxiv.org/abs/1908.09732}{arXiv:1908.09732}]}.

\bibitem{Cheng:2023qys}
Y.-Z.~Cheng, W.-H.~Wu, and Y.~Cao, ``{Electromagnetic Radiation from Binary
  Stars Mediated by Ultralight Scalar}.'' {\ttfamily
  \href{https://arxiv.org/abs/2401.00204}{arXiv:2401.00204}}.

\bibitem{Poisson_Will_2014}
E.~Poisson and C.~M.~Will, {\em Gravity: Newtonian, Post-Newtonian,
  Relativistic}.
\newblock Cambridge University Press, 2014.

\bibitem{Kramer:2021jcw}
M.~Kramer {\em et~al.}, ``{Strong-Field Gravity Tests with the Double
  Pulsar},'' \href{https://dx.doi.org/10.1103/PhysRevX.11.041050}{Phys.\  Rev.\
   X {\bfseries 11} (2021) 041050} {\ttfamily
  [\href{https://arxiv.org/abs/2112.06795}{arXiv:2112.06795}]}.

\bibitem{Weisberg:2016jye}
J.~M.~Weisberg and Y.~Huang, ``{Relativistic Measurements from Timing the
  Binary Pulsar PSR B1913+16},''
  \href{https://dx.doi.org/10.3847/0004-637X/829/1/55}{Astrophys.\  J.\
  {\bfseries 829} (2016) 55} {\ttfamily
  [\href{https://arxiv.org/abs/1606.02744}{arXiv:1606.02744}]}.

\bibitem{vanLeeuwen:2014sca}
J.~van Leeuwen {\em et~al.}, ``{The Binary Companion of Young, Relativistic
  Pulsar J1906+0746},''
  \href{https://dx.doi.org/10.1088/0004-637X/798/2/118}{Astrophys.\  J.\
  {\bfseries 798} (2015) 118} {\ttfamily
  [\href{https://arxiv.org/abs/1411.1518}{arXiv:1411.1518}]}.

\bibitem{Ferdman:2014rna}
R.~D.~Ferdman {\em et~al.}, ``{PSR J1756\ensuremath{-}2251: a pulsar with a
  low-mass neutron star companion},''
  \href{https://dx.doi.org/10.1093/mnras/stu1223}{Mon.\  Not.\  Roy.\  Astron.\
   Soc.\  {\bfseries 443} (2014) 2183--2196} {\ttfamily
  [\href{https://arxiv.org/abs/1406.5507}{arXiv:1406.5507}]}.

\bibitem{Jacoby:2006dy}
B.~A.~Jacoby, {\em et al.}, ``{Measurement of Orbital Decay in the Double
  Neutron Star Binary PSR B2127+11C},''
  \href{https://dx.doi.org/10.1086/505742}{Astrophys.\  J.\  Lett.\  {\bfseries
  644} (2006) L113--L116} {\ttfamily
  [\href{https://arxiv.org/abs/astro-ph/0605375}{astro-ph/0605375}]}.

\bibitem{Fonseca:2014qla}
E.~Fonseca, I.~H.~Stairs, and S.~E.~Thorsett, ``{A Comprehensive Study of
  Relativistic Gravity using PSR B1534+12},''
  \href{https://dx.doi.org/10.1088/0004-637X/787/1/82}{Astrophys.\  J.\
  {\bfseries 787} (2014) 82} {\ttfamily
  [\href{https://arxiv.org/abs/1402.4836}{arXiv:1402.4836}]}.

\bibitem{Stairs:2003eg}
I.~H.~Stairs, ``{Testing general relativity with pulsar timing},''
  \href{https://dx.doi.org/10.12942/lrr-2003-5}{Living Rev.\  Rel.\  {\bfseries
  6} (2003) 5} {\ttfamily
  [\href{https://arxiv.org/abs/astro-ph/0307536}{astro-ph/0307536}]}.

\bibitem{Wex:2014nva}
N.~Wex, ``{Testing Relativistic Gravity with Radio Pulsars}.'' {\ttfamily
  \href{https://arxiv.org/abs/1402.5594}{arXiv:1402.5594}}.

\bibitem{Berti:2015itd}
E.~Berti {\em et~al.}, ``{Testing General Relativity with Present and Future
  Astrophysical Observations},''
  \href{https://dx.doi.org/10.1088/0264-9381/32/24/243001}{Class.\  Quant.\
  Grav.\  {\bfseries 32} (2015) 243001} {\ttfamily
  [\href{https://arxiv.org/abs/1501.07274}{arXiv:1501.07274}]}.

\bibitem{Freire:2024adf}
P.~C.~C.~Freire and N.~Wex, ``{Gravity experiments with radio pulsars},''
  \href{https://dx.doi.org/10.1007/s41114-024-00051-y}{Living Rev.\  Rel.\
  {\bfseries 27} (2024) 5} {\ttfamily
  [\href{https://arxiv.org/abs/2407.16540}{arXiv:2407.16540}]}.

\bibitem{Blas:2019hxz}
D.~Blas, D.~L\'opez~Nacir, and S.~Sibiryakov, ``{Secular effects of ultralight
  dark matter on binary pulsars},''
  \href{https://dx.doi.org/10.1103/PhysRevD.101.063016}{Phys.\  Rev.\  D
  {\bfseries 101} (2020) 063016} {\ttfamily
  [\href{https://arxiv.org/abs/1910.08544}{arXiv:1910.08544}]}.

\bibitem{Kus:2024vpa}
P.~K\r{u}s, D.~L\'opez~Nacir, and F.~R.~Urban, ``{Bayesian sensitivity of
  binary pulsars to ultra-light dark matter}.'' {\ttfamily
  \href{https://arxiv.org/abs/2402.04099}{arXiv:2402.04099}}.

\bibitem{Fabbrichesi:2019ema}
M.~Fabbrichesi and A.~Urbano, ``{Charged neutron stars and observational tests
  of a dark force weaker than gravity},''
  \href{https://dx.doi.org/10.1088/1475-7516/2020/06/007}{JCAP {\bfseries 06}
  (2020) 007} {\ttfamily
  [\href{https://arxiv.org/abs/1902.07914}{arXiv:1902.07914}]}.

\bibitem{LIGOScientific:2017vwq}
{\bfseries LIGO Scientific, Virgo} Collaboration, ``{GW170817: Observation of
  Gravitational Waves from a Binary Neutron Star Inspiral},''
  \href{https://dx.doi.org/10.1103/PhysRevLett.119.161101}{Phys.\  Rev.\
  Lett.\  {\bfseries 119} (2017) 161101} {\ttfamily
  [\href{https://arxiv.org/abs/1710.05832}{arXiv:1710.05832}]}.

\bibitem{Pearson:2018tkr}
J.~M.~Pearson, {\em et al.}, ``{Unified equations of state for cold
  non-accreting neutron stars with Brussels\textendash{}Montreal functionals
  \textendash{} I. Role of symmetry energy},''
  \href{https://dx.doi.org/10.1093/mnras/sty2413}{Mon.\  Not.\  Roy.\  Astron.\
   Soc.\  {\bfseries 481} (2018) 2994--3026} {\ttfamily
  [\href{https://arxiv.org/abs/1903.04981}{arXiv:1903.04981}]}. [Erratum:
  Mon.Not.Roy.Astron.Soc. 486, 768 (2019)].

\bibitem{Hulse:1974eb}
R.~A.~Hulse and J.~H.~Taylor, ``{Discovery of a pulsar in a binary system},''
  \href{https://dx.doi.org/10.1086/181708}{Astrophys.\  J.\  Lett.\  {\bfseries
  195} (1975) L51--L53}.

\bibitem{Blanchet:2013haa}
L.~Blanchet, ``{Gravitational Radiation from Post-Newtonian Sources and
  Inspiralling Compact Binaries},''
  \href{https://dx.doi.org/10.12942/lrr-2014-2}{Living Rev.\  Rel.\  {\bfseries
  17} (2014) 2} {\ttfamily
  [\href{https://arxiv.org/abs/1310.1528}{arXiv:1310.1528}]}.

\bibitem{Will:2018bme}
C.~M.~Will, {\em {Theory and Experiment in Gravitational Physics}}.
\newblock Cambridge University Press, 2018.

\bibitem{Damour:1988mr}
T.~Damour and G.~Schaefer, ``{Higher Order Relativistic Periastron Advances and
  Binary Pulsars},'' \href{https://dx.doi.org/10.1007/BF02828697}{Nuovo Cim.\
  B {\bfseries 101} (1988) 127}.

\bibitem{AIHPA_1986__44_3_263_0}
T.~Damour and N.~Deruelle, ``General relativistic celestial mechanics of binary
  systems. {II.} {The} post-newtonian timing formula,'' Annales de l'I.\ H.\
  P.\  Physique th\'eorique {\bfseries 44} (1986) 263--292.

\bibitem{Damour:1991rd}
T.~Damour and J.~H.~Taylor, ``{Strong field tests of relativistic gravity and
  binary pulsars},'' \href{https://dx.doi.org/10.1103/PhysRevD.45.1840}{Phys.\
  Rev.\  D {\bfseries 45} (1992) 1840--1868}.

\bibitem{LIGOScientific:2024elc}
{\bfseries LIGO Scientific, Virgo,, KAGRA, VIRGO} Collaboration, ``{Observation
  of Gravitational Waves from the Coalescence of a 2.5-4.5 M$_\odot$ Compact
  Object and a Neutron Star},''
  \href{https://dx.doi.org/10.3847/2041-8213/ad5beb}{Astrophys.\  J.\  Lett.\
  {\bfseries 970} (2024) L34} {\ttfamily
  [\href{https://arxiv.org/abs/2404.04248}{arXiv:2404.04248}]}.

\bibitem{Krause:1994ar}
D.~Krause, H.~T.~Kloor, and E.~Fischbach, ``{Multipole radiation from massive
  fields: Application to binary pulsar systems},''
  \href{https://dx.doi.org/10.1103/PhysRevD.49.6892}{Phys.\  Rev.\  D
  {\bfseries 49} (1994) 6892--6906}.

\bibitem{Shapiro:1983du}
S.~L.~Shapiro and S.~A.~Teukolsky,
  \href{https://dx.doi.org/10.1002/9783527617661}{{\em {Black holes, white
  dwarfs, and neutron stars: The physics of compact objects}}}.
\newblock 1983.

\bibitem{Pearson:2020bxz}
J.~M.~Pearson, N.~Chamel, and A.~Y.~Potekhin, ``{Unified equations of state for
  cold nonaccreting neutron stars with Brussels-Montreal functionals. II. Pasta
  phases in semiclassical approximation},''
  \href{https://dx.doi.org/10.1103/PhysRevC.101.015802}{Phys.\  Rev.\  C
  {\bfseries 101} (2020) 015802} {\ttfamily
  [\href{https://arxiv.org/abs/2001.03876}{arXiv:2001.03876}]}.

\bibitem{Liu:2024qzp}
Z.~Liu and Z.-W.~Tang, ``{Probing ultralight isospin-violating mediators at
  GW170817}.'' {\ttfamily
  \href{https://arxiv.org/abs/2402.06209}{arXiv:2402.06209}}.

\bibitem{Tolman:1939jz}
R.~C.~Tolman, ``{Static solutions of Einstein's field equations for spheres of
  fluid},'' \href{https://dx.doi.org/10.1103/PhysRev.55.364}{Phys.\  Rev.\
  {\bfseries 55} (1939) 364--373}.

\bibitem{Oppenheimer:1939ne}
J.~R.~Oppenheimer and G.~M.~Volkoff, ``{On massive neutron cores},''
  \href{https://dx.doi.org/10.1103/PhysRev.55.374}{Phys.\  Rev.\  {\bfseries
  55} (1939) 374--381}.

\bibitem{kittel1980thermal}
C.~Kittel and H.~Kroemer, {\em Thermal Physics}.
\newblock W. H. Freeman, 1980.

\bibitem{Freire:2012mg}
P.~C.~C.~Freire, {\em et al.}, ``{The relativistic pulsar-white dwarf binary
  PSR J1738+0333 II. The most stringent test of scalar-tensor gravity},''
  \href{https://dx.doi.org/10.1111/j.1365-2966.2012.21253.x}{Mon.\  Not.\
  Roy.\  Astron.\  Soc.\  {\bfseries 423} (2012) 3328} {\ttfamily
  [\href{https://arxiv.org/abs/1205.1450}{arXiv:1205.1450}]}.

\bibitem{Antoniadis:2013pzd}
J.~Antoniadis {\em et~al.}, ``{A Massive Pulsar in a Compact Relativistic
  Binary},'' \href{https://dx.doi.org/10.1126/science.1233232}{Science
  {\bfseries 340} (2013) 6131} {\ttfamily
  [\href{https://arxiv.org/abs/1304.6875}{arXiv:1304.6875}]}.

\bibitem{Bhat:2008ck}
N.~D.~R.~Bhat, M.~Bailes, and J.~P.~W.~Verbiest, ``{Gravitational-radiation
  losses from the pulsar-white-dwarf binary PSR J1141-6545},''
  \href{https://dx.doi.org/10.1103/PhysRevD.77.124017}{Phys.\  Rev.\  D
  {\bfseries 77} (2008) 124017} {\ttfamily
  [\href{https://arxiv.org/abs/0804.0956}{arXiv:0804.0956}]}.

\end{thebibliography}\endgroup
\bibliographystyle{utphys28mod}

\end{document}